\documentclass[conference]{IEEEtran}

\usepackage{cite}
\usepackage[pdftex]{graphicx}
\usepackage[cmex10]{amsmath}
\usepackage{algorithmic}
\usepackage{array}
\usepackage[tight,footnotesize]{subfigure}
\usepackage[font=footnotesize]{subfig}
\usepackage{graphicx}
\usepackage{mathtools}
\usepackage{amsmath}
\def\union{{\cup}}

\usepackage{alltt}

\def\boxx{{\vcenter{\vbox{\hrule height.3pt
          \hbox{\vrule width.3pt height6pt
          \kern6pt\vrule width.3pt}\hrule height.3pt}}\;}}

\def\impos{{\;\vcenter{\hbox{\rule{5mm}{0.2mm}}} \vcenter{\hbox{\rule{1.5mm}{1.5mm}}} \;}}

\def\lrarrow{\leftrightarrow \kern-8pt \rightarrow}

\def\beq{\begin{eqnarray}}
\def\eeq{\end{eqnarray}}
\def\2{\frac{1}{2}}

\def\lrarrow{\leftrightarrow \kern-8pt \rightarrow}

\def\frightarrow{\rightarrow \kern-11pt /~~}
\def\reducesto{\simeq \kern -3pt >}

\usepackage{fancyhdr}					
\begin{document}
\newcommand{\strust}[1]{\stackrel{\tau:#1}{\longrightarrow}}
\newcommand{\trust}[1]{\stackrel{#1}{{\rm\bf ~Trusts~}}}
\newcommand{\promise}[1]{\xrightarrow{#1}}
\newcommand{\revpromise}[1]{\xleftarrow{#1} }
\newcommand{\lrpromise}[1]{{\xleftrightarrow{#1}} }
\newcommand{\imposition}[1]{\stackrel{#1}{\impos}}
\newcommand{\scopepromise}[2]{\xrightarrow[#2]{#1}}
\newcommand{\handshake}[1]{\xleftrightarrow{#1} \kern-8pt \xrightarrow{} }
\newcommand{\cpromise}[1]{\stackrel{#1}{\frightarrow}}
\newcommand{\policy}{\stackrel{P}{\equiv}}
\newcommand{\field}[1]{\mathbf{#1}}
\newcommand{\bundle}[1]{\stackrel{#1}{\Longrightarrow}}

\title{A Promise Theory Perspective on Data Networks}

\author{\IEEEauthorblockN{Paul Borrill}
\IEEEauthorblockA{EARTH Computing, Inc\\
Palo Alto, CA 94306\\
paul@borrill.com}
\and
\IEEEauthorblockN{Mark Burgess}
\IEEEauthorblockA{CFEngine, Inc.\\ Mountain View, CA 94040\\
mark.burgess@cfengine.com}
\and 
\IEEEauthorblockN{Todd Craw}
\IEEEauthorblockA{Cumulus Networks\\
Mountain View, CA 94041\\
todd@cumulusnetworks.com}
\and
\IEEEauthorblockN{Mike Dvorkin}
\IEEEauthorblockA{Cisco Inc.\\
San Jose, CA\\
Mike.Dvorkin@cisco.com}
}

\maketitle

\begin{abstract}
  Networking is undergoing a transformation throughout our industry.
  The need for scalable network control and automation shifts the
  focus from hardware driven products with ad hoc control to Software
  Defined Networks. This process is now well underway. In this paper,
  we adopt the perspective of the Promise Theory to examine the
  current and future states of networking technologies. The goal is to
  see beyond specific technologies, topologies and approaches and
  define principles. Promise Theory’s bottom-up modeling
  has been applied to server management for many years and lends
  itself to principles of self-healing, scalability and robustness.
\end{abstract}

\IEEEpeerreviewmaketitle

\section{Introduction} 


As networks grow in scale and complexity, some argue for a return to
centralized and imperative management\cite{Levin:2012:LCS:2342441.2342443}.
Network design revolves around legacy data structures and protocols rather
than the business functionality required from the network. A modern
approach to network design emphasizing simplicity and relevant
abstraction seems overdue.  Such an approach could reduce the cost and
brittleness of network design.

The `Promise Theory', was introduced in 2005 as a way to model
distributed systems with complete
decentralization\cite{burgessdsom2005}.  Coupled with abstraction, it
offers a looking glass onto the design and management of networks. If
we define what a user or application needs from the network we can
begin to get away from imperatively controlling the `how' the network
functions and instead focus on declaratively describing "what" is
required from it. In Promise Theory, network elements act as autonomous agents and
collaborate to find the best way to deliver the required function.

In this paper we apply Promise Theory as a measuring stick for
the current state of networking to cast a critical eye over current
practice and future directions. We show that there are simple
unifying principles for networking that are independent of scaling
arguments, and that there is no need to base future networking on
centralized control.

\section{Promise Theory}
                                
Promise theory is about what can happen in a collection of components
that work together\cite{burgessdsom2005,mark_burgess_voluntary_2005}.
It is not a network protocol, but a descriptive algebra.  One begins
with the idea of completely autonomous agents that interact through
the promises they make to one another. It is well-suited to modeling
networks \cite{m._burgess_scalability_2004}. Although we cannot force
autonomous agents to work together, we can observe when there are
sufficient promises made to conclude that they are indeed cooperating
voluntarily.  Our challenge in this paper, is to
translate this bottom-up view into top-down, human managed
requirements.

\emph{Agent} is the term used for the fundamental entities in Promise
Theory.  Agents are not necessarily like `software agents', they can
be any active entities like an interface that keeps promises.  Actions
taken by agents are not in the scope of Promise Theory. We assume that
appropriate actions are taken to keep the promises. In that way, we
focus on declarative intent, rather than imperative procedures.

\subsection{Formalism}

The promise formalism has a number of features, described in\cite{bergstra_promise_2014}.
We refer readers to this reference for details. 

A {\em promise} is an intention that has been `voluntarily' adopted by
an agent (usually channeling a human owner, or perhaps an agreed
standardization).  An agent that only promises to do as it's told is
{\em dependent} or voluntarily subordinated.  It has some of the
characteristics of a service: an agent makes its intended behavior
known to other agents (e.g. I will serve files on demand, or forward
packets when I receive them).  An {\rm imposition} is an attempt to
induce the cooperation of another agent by imposing upon it (e.g. give
me the file, take this packet).

We write a promise from $\rm Promiser$
to $\rm Promisee$, with body $b$ as follows:
\beq
{\rm Promiser} \promise{~b~} {\rm Promisee}.\nonumber
\eeq
and we denote an imposition by
\beq
{\rm Imposer} \imposition{b} {\rm Imposee}.\nonumber
\eeq
Promises and impositions fall into two polarities, denoted by
$\pm$. A promise to give or provide a behavior $b$ is denoted by
a body $+b$; a promise to accept something is denoted
$-b$ (or sometimes $U(b)$, meaning use-$b$). Similarly,
an imposition on an agent to give something would have body $+b$, while
an imposition to accept something has a body $-b$.

Although promises are not a network protocol, agents can exchange data.
To complete any kind of exchange, we need a match an imposition (+)
with a promise to use (-).  To form a binding (as part of a contract),
we need to match a promise to give (+) with a promise to use (-).
This rule forces one to document necessary and sufficient conditions
for cooperative behaviour.

A promise model thus consists of a graph of nodes ({\em agents}), and
edges (either {\em promises} or {\em impositions}) used to communicate
intentions. Whatever protocol might be used to communicate promises is
not defined (and shouldn't be). Agents publish their intentions and
other agents may or may not choose to pay attention.  In that sense,
it forms a chemistry of intent \cite{burgess_search_2013}, with no
particular manifesto, other than to decompose systems into the set of
necessary and sufficient promises to model intended behavior.

\section{Networking from a promise perspective}

\subsection{Ethernet (L2)}

The agents that keep promises to send and receive data are the network
interfaces. For example,
in the Ethernet protocol, interfaces $E_i$ promise to label transmissions
with a unique MAC addresses or string of digits.  
\beq
E_i &\promise{+{\rm MAC}_i | {\rm MAC}_i \not= {\rm MAC}_j}& E_j ~~~\forall i,j\nonumber
\eeq 
When data are transmitted by an interface, the interface keeps its promises to use
messages that have (destination MAC address, data).  Note: the message
is not a promise, the promise governs how the message is handled.
\beq E_i
&\imposition{(+{\rm MAC}_j, +{\rm data})}& E_j \nonumber
\eeq 
Messages are sent
`fire and forget' as impositions on to a remote receiver.  While all
interfaces generally promise to accept any MAC address, (unless they
block with MAC access control) only the interface whose MAC address
matches the destination in the message doublet actually promises to
accept the message voluntarily. Note, there is nothing
other than convention to prevent all agents from accepting the data
too; this `promiscuous mode' is used for network monitoring,
for example.  
\beq
E_* &\promise{-{\rm MAC}_j}& E_i ~~~\forall i,j\nonumber\\
E_i &\promise{(-{\rm MAC}_j, -{\rm data}) ~ {\rm if}~ (i=j)} & E_j\nonumber
\eeq 
Since the channel is unprotected, agents effectively promise the
data to all others in scope.  Moreover, all agents promise to decode
the address and the data, but many will discard the results.

While this set of promises is scale independent, the assumption that
every agent has to be in scope of every transmission does not scale,
since it requires messages to be flooded or broadcast to every node (agent), in
principle.  The primary issue with raw Ethernet is that there are no
ways to selectively limit the size of these broadcast
domains.  This makes the `everyone please take a look at this'
approach impractical.

\begin{figure}[ht]
\begin{center}
\includegraphics[width=4cm]{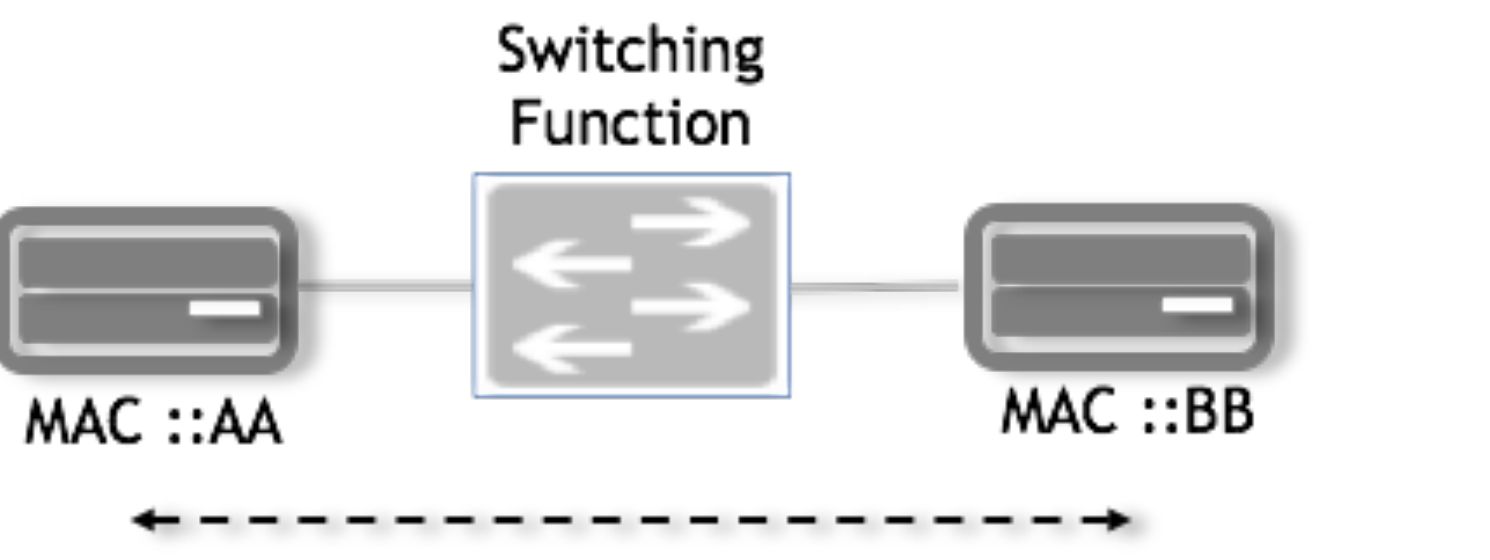}
\caption{\small An Ethernet switching function. \label{switch}}
\end{center}
\end{figure}

In Fig. \ref{switch} we see two interfaces that promise MAC address
00:00:11:11:11:AA (shortened to AA) and 00:00:11:11:11:BB (shortened
to BB).  Suppose we wish to send data from AA to BB, then, since the
Ethernet is a push-based imposition protocol, only half a contract is
needed for emergent delivery, and we leave the rest to
trust.

\small
\beq
E_{\rm AA} &\imposition{\rm +MAC_{\rm BB}}& E_{\rm switch}\nonumber\\
E_{\rm switch}&\promise{\rm -MAC_{\rm i}}& E_{\rm i} ~~~ \forall i\nonumber\\
E_{\rm switch} &\promise{\rm  +forward~MAC~_{\rm BB}}& E_{\rm BB}\nonumber
\eeq
\normalsize

In each point-to-point interaction, the agent has to formally promise
to use (-) the delivery service promised by the agent giving (+). This
is the algebra of binding. There is no notion of a permanent virtual
circuit, as say in ATM.  However, if we add handshaking, a similar
story can be told about ATM, Frame Relay, MPLS and other systems.

\subsection{Internet Protocol (L3)} \label{ipsec}

IP provides Wide Area Networking by issuing two part addressing to
cope with transmission scalability. IP addresses still promise to be
globally unique, but are interpreted as doublets.
\begin{quote}
(network prefix, local address)
\end{quote}
Only addresses with the same prefix are considered in mutual scope
for broadcasting, and messages addressed from one prefix to another
promise to be forwarded deliberately rather than
by `flooding'. IP is thus a cooperative effort that builds on promises
rather than impositions alone.

\begin{figure}[ht]
\begin{center}
\includegraphics[width=7cm]{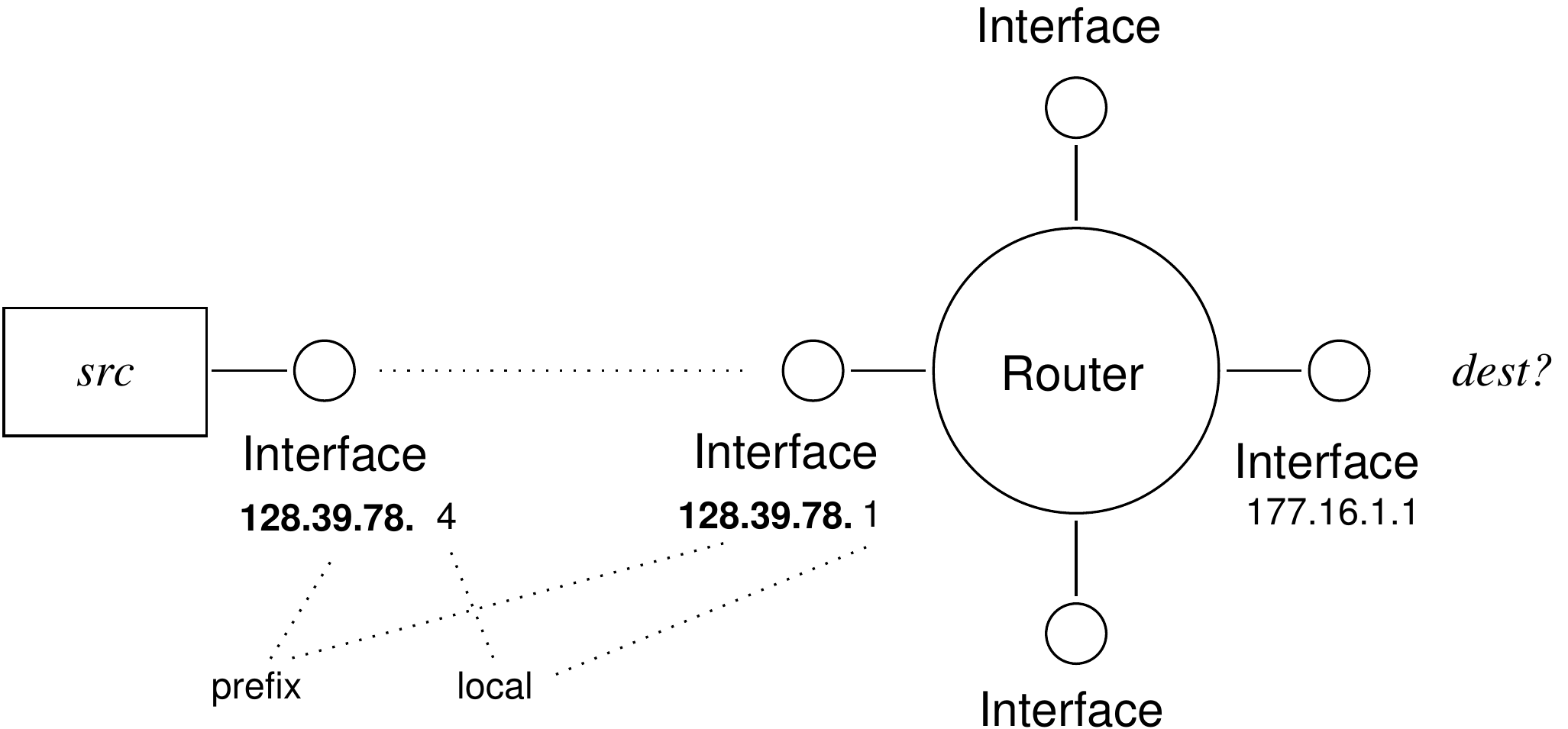}
\caption{\small Internet promises. An end-node or leaf and its
single interface promises to relay through a `router'
which is surrounded by multiple interfaces, thus connecting
multiple network branches.\label{iproute}}
\end{center}
\end{figure}

To make this work, IP needs two kinds of agent, which fall into
different promise roles (see figure \ref{iproute}): {\em interfaces}
(terminating connections), which only transmit and receive data
intended for them, and {\em forwarders} (called routers or switches)
that cooperate with multiple interfaces, and promise to selectively
forward data from one interface to another between protected
broadcast domain.  This acts as a flood-barrier or firewall to packets
promised to different prefixed networks.

To model routers, without giving up the interface abstraction,
we introduce the concept of a route service (or link service),
whose job it is to establish cooperative forwarding between the
interfaces.

Consider Fig. \ref{iproute}. The source node has an
address, normally written 128.39.78.4/24. As a doublet, the promises
see it in two parts as $i = $ (prefix=128.39.78, local=4). We'll call this
the source prefix, or, 
$j = $ (prefix=128.39.78, local=1) for the router interface. 
When a message is sent to an address with a
different destination prefix, data are sent by imposition to the
interface on the router with the source network prefix (usually the
`default route'):
\beq
I_{source_i} \imposition{+({\rm destination, local}), +{\rm data}} I_{\rm router_j}\nonumber
\eeq
Each router interface $j$ promises the connected source interfaces $i$ to use
all such packets, a priori, and to present them to the router (kernel)
which keeps the following promises.
\beq
I_{\rm router_j} \promise{-(*, *), -{\rm data}} I_{{\rm source}_i}\nonumber\\
I_{\rm router_j} \promise{+{\rm prefix}, +{\rm data}} {\rm Router}\nonumber
\eeq
Similarly, other interfaces connected to the router's interfaces
promise to accept messages from the router that have their prefix:
\beq
 I_{\rm source_i} \promise{-({\rm prefix, source}), +{\rm data}} {\rm Router_j}\nonumber
\eeq
Crucially for messages to escape from a local region, the router
promises all IP interfaces to forward messages it receives 
on one if its own interfaces according to a set of promises
which we denote `forward'. The router interfaces, in turn, bind to this
promise by accepting it.
\beq
{\rm Router} \promise{+{\rm forward}} I_{\rm router_j}\nonumber \\
I_{\rm router_j} \promise{-{\rm forward}} {\rm Router}\nonumber
\eeq
The forward promise has the following logic:
\begin{quote}
\small(1) If the prefix of the destination interface is the same as the prefix
of one of the router's interfaces, forward the message onto that interface.
\end{quote}
The remainder of the promise requires configuration with knowledge
of the wider world.
\begin{quote}
\small (2) If the prefix of the destination interface is known to an internal
database of external knowledge, i.e. the
Routing Information Base (RIB),
forward the message to the interface
known to lead to the desired destination.\\
~\\
(3) Send all other message destinations to a pre-decided default interface,
where we expect to reach some other router with greater knowledge
of how to find the prefixed network.
\end{quote}

Note that, like the Ethernet, this algorithm has only emergent
behaviour that matches its design goal. It cannot, by direct
imposition, assure a successful delivery of messages, because that
requires the cooperation of potentially many intermediate interfaces
and routing agents. In spite of this apparent lack of control, the Internet
works demonstrably well.
Trust plays a major role in operations. 

\subsection{VLAN: L2 channel containment}

The concept of doublet addressing in IP enabled improved scalability,
by black-boxing local networks, but added the cost of routing. 
How expensive routing is, in relative terms, is a constantly changing
overhead that depends on many current technological factors. Fear of
this cost tends to make datacenter traffic favor L2 solutions.

Routers were optimized for WAN delivery, so the obvious question for
LAN managers was: could routing be simplified for smaller local regions
without the paraphernalia needed for global routing? 

When packets don't have to be routed through multiple hops, i.e. when
parts (2) and (3) of the forwarding promise can be ignored, a simpler
form of prefixing can be used. This is the concept of the VLAN
overlay.  Interfaces can simply be classified, or tagged with short
integer labels:
\begin{quote}
(prefix, local address) $\rightarrow$ (VLAN-id, MAC-address)
\end{quote}
Then we have multiplet address components again (but now with a short VLAN tag
instead of a large integer prefix), for boxing
off local regions. A VLAN tag signifies membership in a private logical
container. As with Frame Relay, these tags have to be configured
manually, so routing is a human-centric process.

The concept of a Level 2 overlay has become quite popular for its
perceived simplicity in small isolated networks. It's limitations have
to do with scaling of the manual configuration and broadcast domains.
VLAN is a brute force routing mechanism that scales linearly with the
number of addresses in a container. Containers are not localized in
physical space, only in logical channel space (unlike the assumed
distribution of prefixes in IP). Thus this does not address the issues
of physical scaling.  However, we need something that scales like
$\log N$ or better (like IP). We return to this in section
\ref{scale}.

\subsection{Tunnelling addresses and transducer pattern}

Embedding protocols inside one another is not the only approach to
containment. One can also strip off and repackage data on different
legs of a journey. To do this, one makes a transducer that converts
one kind of addressing into another (see \cite{bergstra_promise_2014}).

ARP is one such service that maps between Ethernet MAC addresses and
IP addresses. Instead of a physical forwarding table, a logical rewriting table
is maintained.  When a direct ARP conversion is not possible, data are
sent to the default route, which is the address of the router
interface $I_{\rm prefix}$ to the default interface.
DNS is another transducer, that maps from symbolic addresses
to IP addresses. 

The same principle has been applied to isolated networks, such as the
reserved namespaces 10.0.0.0 and the example.com addresses
192.1.168.0/24.  IP Network Address Translation (NAT) is now been
promoted from crude workaround to viable technology, extending the
local IP addressing component with additional internal addressing
numbers, the rewriting outgoing addresses to point to the standard IP
address range.  End to end addressability is not normally promised in
this scheme however, so it has limited value, (however see
TRIAD\cite{triad}) for a viable scheme for extending IPv4 in this manner.

More recently, a tunnelling approach is also being used to
artificially extend Layer 2 VLAN as a stop-gap measure for a
technology users who are familiar with VLAN. 
VxLAN, and NVGRE are encapsulations of Ethernet L2 Frames, with
tunnelling over IP to enable the physical reach across multiple
gateways.  Addresses add a multiplet component: a Tenant Network
Identifier (TNI) or Virtual Tunnel End Point (VTEP) identifier
embedded parallel channels.  

These schemes perform two functions: i) they increase the number of
possible VLAN-like channel addresses, patching a limitation in the VLAN
implementation, and ii) they allow teleportation of broadcast domains
across an IP scale network, transparently of routing concerns.
Thus they do not eliminate the cost of IP routing, but offer a comfortable
user interface for local network administrators.


\section{Promise principles for scalable networking}

Let us consider the principles that summarize the work in the previous sections.
We see two main issues: container-specific addressing, and interface
to interface forwarding.  Multiplet addresses allow the containment of
broadcasts as well as selective routing of traffic by `prefix'.

\subsection{Addressability with scope or namespaces}

By introducing multiplet addressing, we draw a logical (and perhaps
physical) container around a network region which hides its internals
with some kind of identifier or prefix, which acts as a namespace
identifier. Everything inside the namespace is local and protected.
There are two principles that explain these cases.

\begin{quote}{\em Principle 1: Container multiplet addressing}: 
\small Any system that promises to support $n$-tuple addressability of parts, for $n > 1$, 
enables logical or physical containment of information, as well as
log-scalable routability between the containers.
$\bigtriangleup$
\end{quote}

To transmit data across multiple (possibly embedded) containers, we typically need an address
component for each logical container. Thus interfaces $a_i$ must promise to recognize
one of the components $a_i$ and pass on all others as passenger data:
\beq
{\rm Interface_i} \promise{\pm(a_1,a_2,{\bf a_i},\ldots,a_n),\pm {\rm data}} {\rm Router}\nonumber
\eeq
e.g. the $a_i$ address components might include MAC address, IP address, VLAN number and VxLAN IDs.
This set of addresses need to be configured and managed, either manually or by
some mapping service. Some of these addresses overlap (like IP-LAN and MAC addresses).

\begin{quote}{\em Principle 2: Forwarding by multiplet address}:
\small  Forwarding of multiplet addressed data requires an infrastructure of
  forwarding promises by each members of each container for each
  address component in which all other components are ignored by other
  containers as payload data.  $\bigtriangleup$
\end{quote}

An interfaces $a_i$ in a given container of level $i$ would promise to
accept other components addresses components as data only to be
forwarded, not interpreted, i.e. as payload with no assumed semantics.
In practice some of the address components might be removed or even
rewritten, depending on the encoding as data traverse container
boundaries, but that is not a requirement of the principle. 
All of the components have a continuing logical existence. It would
be enough to ignore them.  Note also that intrusion
detection/prevention systems sometimes break the semantics of ignoring
payload.

\subsection{Addressable scalability}

The scaling of multiplet addressing is a straightforward idea.
It prevents a local namespace from becoming too large for
flooding or broadcasting. The size of the namespace is limited either
by a fixed number of nodes accessible in one multiplet address (e.g.
VLAN tag, OSPF areas, BGP AS, etc), or equivalently, by the size of a prefix in a binary
encoding of the multiplet (as in IP). In the first case, there is no
defined limit to how many MAC addresses can occupy the same VLAN. Scaling
is throttled by physical limitations. In the latter case there is an explicit
quota tradeoff between local and global from a fixed number of addresses.

If an $n$ bit address has a prefix of length $p$, this improves
scaling through black-boxification of local regions. It transforms the
addressing of $N = 2^n$ things into the addressing of merely
$N_C=2^p$, things globally and $n_C= 2^{(n-p)}$ things inside each of
the $C$ containers.  That is $\log_{n_C}$ rather than $n_C$ scaling.

There is also no particular reason why IP addressing has to be limited to
prefix quotas.  IP Network Address Translation is an attempt to extend
the range of local addressing, independently of the prefix quota space
to alleviate IPv4 address depletion.

\subsection{Addressable multi-tenancy}

As a side-effect of supporting logical or physical containment one
obtains the ability to support {\em multi-tenancy}. This can also be
considered a mode of scaling in which one assigns containers to
distinct organizational owners. This requires a mapping service (like an ARP
table for organizations).  Registration of tenants is a manual human
process. Presently there is IANA as a global directory service, ISPs
for address delegation, and the Internet Exchanges for registering
tenancy promises.

\subsection{Two distinct network services}

Networking supports two main use-cases:
\begin{itemize}\small
\item {\em Content delivery} or pull requesting (asynchronous retrieval promises of the form Node $\promise{X}$ Node).

\item {\em Signalling} or push notification (synchronous impositions of the form Node $\imposition{X}$ Node).
\end{itemize}
The former is a many-to-one association, for which we can employ
versioning, replication (data-model de-normalization), re-direction,
and delocalization (e.g.  Content Delivery Networks). Point to point
addressing is less important; caching is highly meaningful. The
concept of Name Based Routing has been proposed to abstract away
endpoint addresses\cite{BariCABM12}.

For the signalling, we still need endpoint-resolution addressability, as
signals cannot be cached, though they might need to be flooded.
Service delivery generally involved a mixture of these two cases,
which depends on the nature of the application being supported.
Applications typically want to make certain promises about
connectivity, security, e.g. load balancing and firewall filtering
options

Application-oriented delivery suggests other forms of containment
based on the logic of the service interaction. Current networking
management abstractions make application specific requirements painful
to configure because of lack of a consistent model for abstracting them.
This brings us to the present day.

\section{Generalizing network containers}

Given that the principle of containment is so flexible and important
for scoping data communications, it is natural to ask whether there
are other more relevant abstractions that better support the needs of
users.  Several authors have opted to rethink network
architecture\cite{balakrishnan,BariCABM12}. We consider this from a
promise theory perspective.

\subsection{Software Defined Networking} 

Software defined networking (SDN) is an umbrella term for a
programmatic approach to managing network devices, using software
controls to replace manual
configuration\cite{Raghavan:2012:SIA:2390231.2390239}.  It is
considered more wide-ranging than the NETCONF or SNMP device
management protocols of the past.
SDN includes concepts related to network virtualization through
overlays, network function virtualization (NFV), virtualized L4/7 service insertion
and control, management and orchestration of physical and
virtual networking devices and functions (such as switches, routers),
as well as control systems relying on specific control protocols like
OpenFlow and OVSDB.

SDN is an interface transducer that essentially virtualizes
existing concepts such as virtual end-points (computers), virtual
cables (L1), virtual switches (L2), etc.  Initial motivations
for SDN were to overcome the brittleness and the lack of
programmability, manageability and agility in networks.
The inability to treat network infrastructure uniformly gave birth to
the network virtualization through overlays, where underlying physical
underlays are abstracted away in a way that provides perceived
uniformity to the application/tenant traffic.

With abstractions rooted in past technology choices, designed for the
pleasure of network engineers rather than application designers, a
cultural gap has been exposed between the needs of application
programmers, writing increasingly distributed software.  The current
networking model requires decomposition into primitives like L2
networks (VLAN, VXLAN), tenant L3 domains (VRF) and multiple rules
expressed as Access Control Lists (ACLs), or, in case of pure
OpenFlow-based SDN, a set of flow rules.

Application architects think about application components and
component interactivity; they rarely think about networks, firewalls
and other L4-7 services. Instead, they think
about services providing \emph{functions} to other services, while
consuming functions of the infrastructure and other application
components/services.

With the emphasis on programmability of legacy technology,
the business purpose of an application can quickly be lost in low
level details. This makes application design and updates a painful
process.

\subsection{Application-Oriented Network Promises} 

Another approach is to rethink the way architects reason about
interaction with the network altogether.  Application-centric
group-based policies could be used for consistent enforcement of
service requirements.  Such `requirements' (impositions) can also be
turned around as promises, specified by the application architect.
This makes a concise self-documentation of purpose, so it also
has positive semantic value. 

Promise Theory provides a theoretical framework for such policy
abstractions too, and so can be instrumental in solving a larger problem:
how to build a scalable, self-stabilizing network, supporting any kind
of abstraction.

There is no particular reason why the container principle cannot be
applied to other logical elements than channels and network
boundaries.  A piece of software can also be viewed as a local network
that connects to a wider area.  Application architects need to think
of network behaviors as application components (the application
services) and manage their interactivity.  Application architects can
then focus on understanding what functions a given service provides
and relies on through exposed interfaces.


Such an application service might be thought of as a cell or logical
network container, encapsulating a number of application servers that
provide the service contained within. All server end-points within a
service promise a set of functions to the consumers of the service.
The collection of such (+) promises can be thought of as the basis for
a contract by which others interact with this service.  In a cell
analogy, this could be called the cell membrane. The cell membrane
protects and regulates the conversations in and out of the cell.  The
(-) promises that listen for client impositions could be thought of as
`receptors' that identify what services can be talked about and how.

Consider a classical three-tier application made from cells 
(promise agents clusters):
\begin{itemize}\small
	\item WEB - Web services.
	\item APP - Application services (e.g. Tomcat or jboss)
	\item DB - database services.
\end{itemize}

The tiers have internal structure, but insofar as we only interact with
their promises, we don't care what it is.  Each tier comprises
multiple computational end-points or service `hosts'. A host or
end-point might be a Virtual Machine (VMs), a container (like LxC or
Solaris Zones) or simply a bare-metal compute instance with a network
connection. Since all hosts within a tier provide identical function
within the application, they form a `promise role' group that makes
identical security, forwarding and QoS promises.  Membership of these
end-points within a group/role can be either administratively or
automatically established, most commonly based on the properties of
the compute element (e.g. virtual machine attributes).

The infrastructure promises made between these tier-agents have to be established.
To avoid the kind of manual setup problems of ATM, VLAN, CLI, etc, 
the end-user's needs are communicated by some kind of signalling impositions (API calls), e.g.
`I need the db to talk to the app and the firewall needs to be opened,
I want sufficient channel capacity (``bandwidth'') to be promised'.
As long as the service accepts such impositions from clients, it would try to
promise what was required. Only the service has the necessary knowledge
about whether such a promise is plausible however, so the decision belongs
inside as an autonomous decision.
 
Each tier is really a collaborative group of agents that not only
promises to provide and use services between one another, but which
internally promises to collaborate uniformly to stay coordinated.

\begin{figure}[ht]
\begin{center}
\includegraphics[width=8cm]{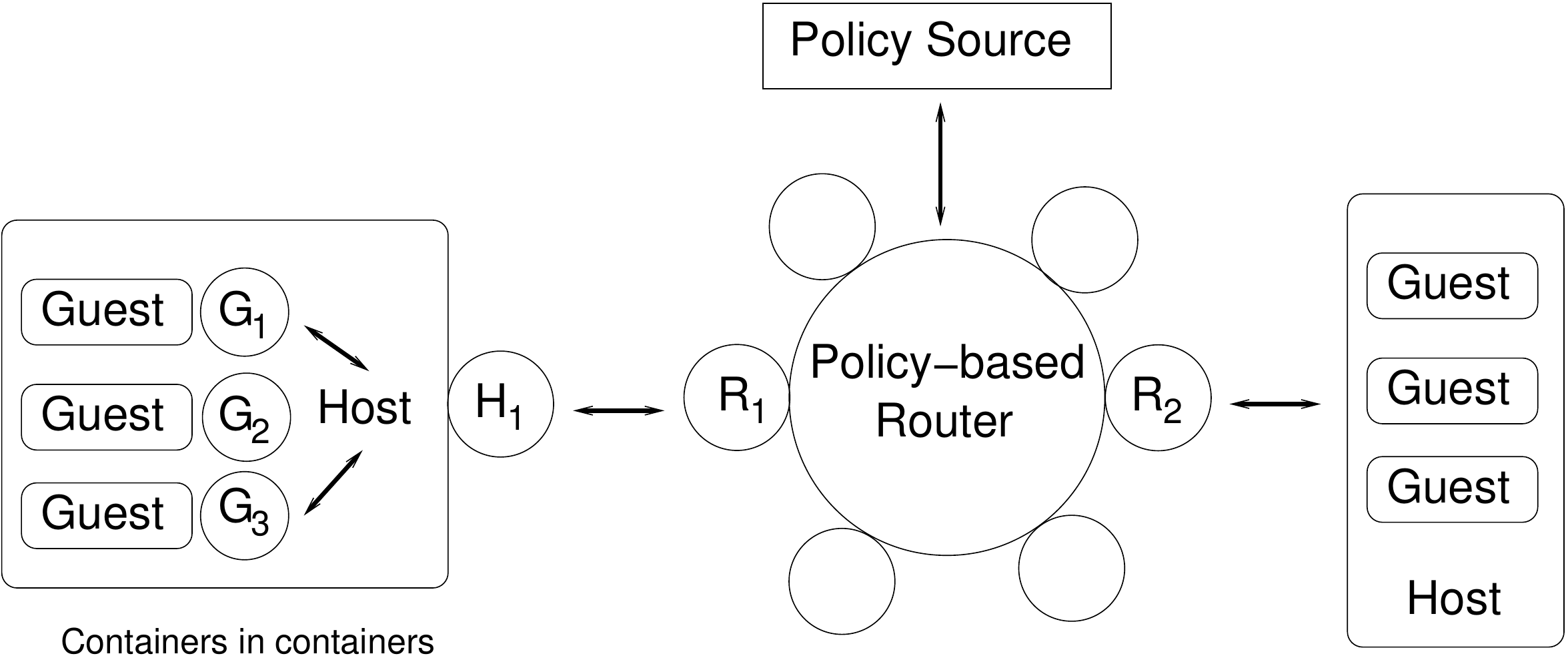}
\caption{\small Host containers promise to forward
communications between guests and `outside' as `routers'. Two levels of container means agents 
need to maintain triplet addresses (Guest, Host, Outside)\label{containers}.
All controllers are embedded in the containers.}
\end{center}
\end{figure}

The challenge of this abstraction is: how do we create a unified user
experience around this for application engineers?  
Users need to comprehend a design composed of an increasing number of
logical entities and layers, arising from cross-cutting abstractions.
We believe the answer here lies in the managing patterns of promises,
which can also be described in terms of the container principle.  The
challenge for technologists is to avoid the temptation to push
complexity back onto users. Hopefully one can avoid referring to the OSI
layer technologies altogether and move towards a description based on
Service Level Agreements (SLA). Current SDN makes this accessible but
not natural. The key would seem to be some kind of
promise compiler.

\subsection{End-to-end service promises via proxy}

To see how this could be done, we return to promise theory.  The
`proxy' or intermediate agent pattern was described in
\cite{bergstra_promise_2014} abstracts the end-to-end delivery promise
of a service $S$ through some promise to handle the delivery details
by proxy $P$ to a client. Both client and server may be
guests running inside various containers.

\begin{figure}[ht]
\begin{center}
\includegraphics[width=4cm]{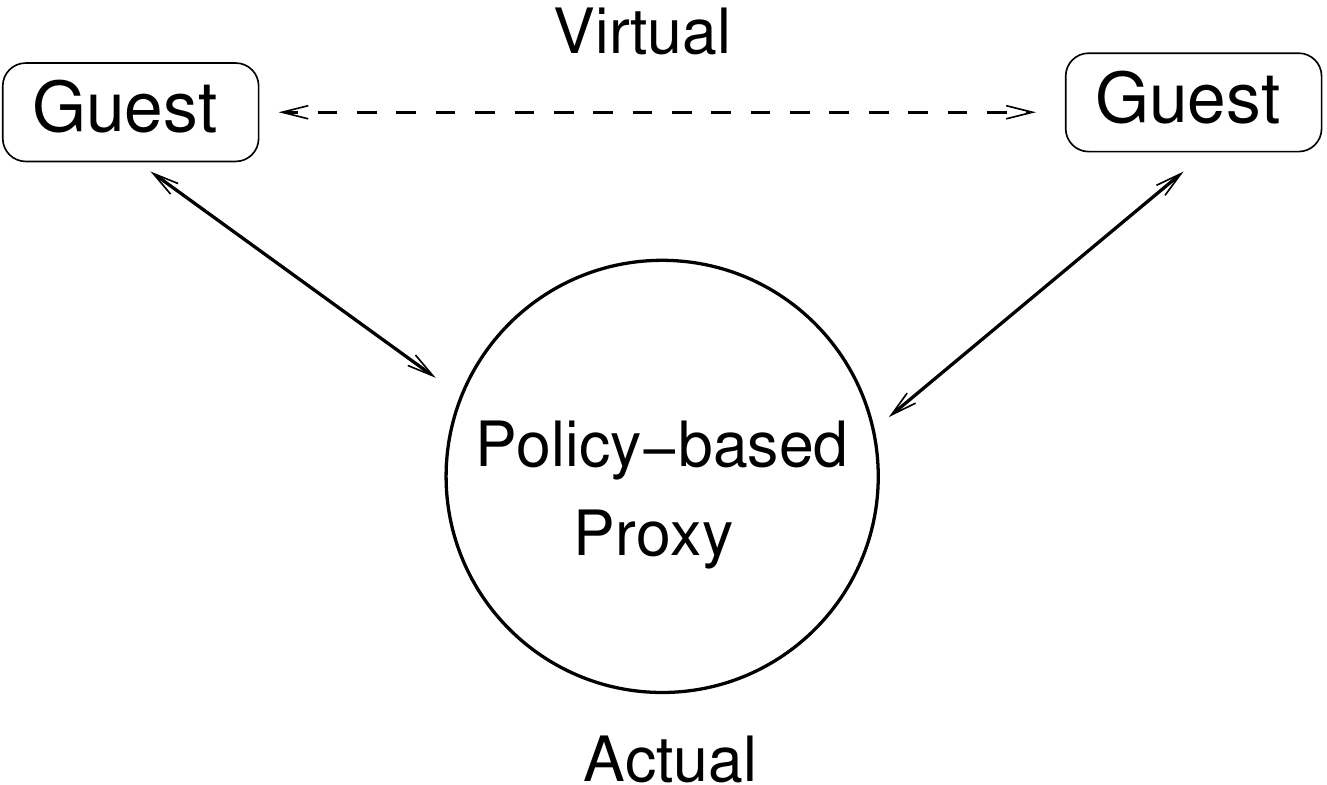}
\caption{\small A generic proxy model that promises to mediate communication, with no internal details exposed.\label{proxy}.}
\end{center}
\end{figure}

The abstraction we would like to expose to the user is for logical services and consumers to
simply make promises directly to one another (Fig.\ref{proxy}), without worrying about all the intermediate
agents in between. The proxy pattern shows how this can be achieved.
We refer readers to \cite{bergstra_promise_2014} (section 11.3) for
a discussion.

Examples of the service $S$ could be: to provide connectivity over a
secure channel, to grant or deny access to data, to commit to or
retrieve from storage, to provide web transport.  It is important to
note that a go-between might create a superficial similarity of function,
but it also adds four promises and hence four possible points of failure
to the equation.

In terms of the cell membrane analogy described above, we would say that the
outer-membrane promises identify how others can communicate to that
service, what they can communicate about, and what happens to the
traffic when they communicate.  Today, one would
have to separate it into VLANS, Firewall rules, Load-Balancing
rules etc. These are details we would prefer to leave to a proxy
on imposing on it a minimum of application specific requirements.  

By providing this as a service, one is able to reason about
applications, and how they interact with other applications rather
than with the physical underlay.  The membrane thus provides a level of
abstraction that hides the details of the cell composition. This is
how the Insieme architecture works.

By isolating promises as containers that promise to play certain roles in an application
design,  one can think about the datacenter as an organism. Then an organism
comprises of many boxes containing multiple functions that manage their own resources
based on a policy declaration.


\subsection{Policy-based containment}

Putting the pieces together, what one ends up with from a promise
perspective is a set of logical containers that keep user-friendly
promises and conceal engineering details. This is not an imposition
control system, but rather a scaled design that compiles into continuously
enforcable local low-level configuration promises.  Based on net-wide
coordinated policy, one compiles high level promises into a kind of
assembler code for low level network configurations supporting a
business purpose.

\section{Fitness for purpose}

We have shown how to represent both low and high level intentions in
terms of promises made by autonomous entities (agents) in IT networks.
We have also indicated how a distributed application can itself be
viewed as a service-oriented network built from autonomous components.
The container principle serves both to abstract and limit the effect
of irrelevant load on inappropriate parts of such collectives.
Containers, not merely as abstract layers, but as namespace boundaries
for autonomous agents to work in, are clearly a key abstraction for
scalable service-oriented data communications.

The next obvious question to ask is: is a set of network containers
{\em fit for purpose}?  How indeed might we align an abstracted networked system
with a given business purpose?

A container (a promise super-agent) can be identified by the pattern of
promises that are made to other agents beyond its `border' or
`membrane'. This promise pattern becomes associated with the {\em
  role} of the container\cite{bergstra_promise_2014}, just as any
namespace label can act as a role identifier. Containers
that play similar roles make functionally equivalent promises.

Promise Theory answers the question of alignment simply in terms of these outward
intentions.  Suppose we can compile our desires for an application or goal into a
number of promises with bodies $d_1, d_2, d_3\ldots$ that would need to be kept.
Then, if our network actually makes promises:
\beq
{\rm Application Network} \promise{b_1, b_2, b_3\ldots} {\rm Users}\nonumber
\eeq
Then we can say, straightforwardly, that the network is aligned with the purpose if
\beq
d_1 \union d_2 \union d_3 \ldots = b_1 \union b_2 \union b_3 \ldots\nonumber
\eeq
The outward promises form the user interface to the containers. Clearly
application oriented containers are more directly business-purpose friendly 
than OSI-layer containers for L2 or L3 abstractions.
But there is more to business purpose than just functionality.
Other systemic promises need to be considered too when discussing
fitness for purpose.  Will the network architecture scale? Will it
fail gracefully under pressure? Will it respond to our needs?
Many of these answers follow from the graph-theoretic properties of the promise graph.
Let's consider these in briefly turn.

\subsection{Distributed control}

Control is the ability to exert influence over the parts of a system.
Traditionally, one looks for:
\begin{itemize}\small
\item A single point of remote control, or
\item A single point of policy calibration.
\end{itemize}
In promise theory, control can only come from within an agent
as an autonomous decision to promise something; for instance,
a promise to provide a service $+S$, such as a function or behaviour,
or a promise to accept a service $-S$, like an Access Control List promise.

Since an autonomous agent is free to ignore signals and impositions
from outside, remote control requires an explicit promise by all
agents to accept impositions from a single point of command. This is
fragile and it is a serial signalling regime.  A more efficient
approach is to turn the signalling problem into a parallel
content-delivery problem by distributing pre-decided policy that
distributed agents can cache for self-correction. Thus, Promise
Theory suggests that a policy model is an improvement in reliability,
fault tolerance and scalability.

With a central controller approach, the central entity has to make
regular impositions on different parts of the system. This adds
complexity from outside the system and makes the controller point
location a bottleneck. A promise based approach allows self-control
based on a cached policy, without further signalling.  Thus it avoids
to cost and resilience issues of signalling infrastructure.  Since
policy is a set of pre-decided promises with associated context which
has potentially global scope, it does not have to be communicated in
real time, it can be cached and replicated efficiently.

\subsection{Resilience and fitness for purpose}

Redundancy plays two roles: one in scaling, for parallel throughput of
data, and another in failover resilience (structural  plasticity) of agents\cite{burgess_search_2013}.
Traditionally one thinks of:
\begin{itemize}\small
\item Single points of failure (unique fragile point).
\item Single point of contact (redundancy allowed).
\end{itemize}
Redundancy means extending
service resources inside a service container in a fashion that is
transparent to a client. 

In promise theory, the generative rule is that no agent may promise
anything that is not about its own behaviour. Thus resilience comes
 `from within'.  Each agent promises its own responsibilities
within the whole. For $+S$ promises each agent can only do its best
and presume additional support from others it knows nothing about.
For $-S$ promises to consume a service agents must say something like
`I promise to use the service $S$ from either agent $X$ or $Y$ or $Z$'
(not just X).  The promise principle forces the responsibility for
resilience back onto the end-point, i.e.  client or server, to the
edges of the network. No middle boxes like load balancers can improve
the situation, because no agent can influence their behaviour.

Fault tolerance can be understood as having clusters of hosts
working cooperatively as `super-agents' to
keep the same forwarding promise. Each agent in
a redundancy group promises to cooperate with every other with mutual
cooperation promises Each agent promises to try
to failover to other agents; thus load balancing also becomes
a client function (without intermediaries). These promises are typically roles by
association\cite{bergstra_promise_2014} (e.g. by policy coordination) rather
than physical data links. The Clos networks 
are realizations of these, in varying degrees of approximation\cite{al-fares_scalable_2008}.

This pattern can be applied both physically and logically between any
pair of roles: client/server, interface/switch, network/router, etc.
For a network, the intra-agent promises inside each super-agent do not
have to be physical connections, but rather intentions to act
symmetrically.  Note however, the architecture reported by Facebook to
formalize this\cite{facebook}.


If the goal of containment is to reduce complexity, then this
architecture may be promised by a simple pattern.  If the promises
made for communication are simple, then the complexity of
applications can also be eliminated without abstraction, by employing
simple stock patterns, like this or like a lattice. For example using
a simple configuration engine to enforce the distributed pattern.
Complex management interfaces with extensive programmability become
unnecessary.

\subsection{Network scalability}\label{scale}

Scalability of  networks is a complex issue that describes how a system can support
increased usage, component addressing, data throughput, without
increased latency.
\begin{itemize}\small
\item Physical connectivity needs sufficient serial/parallel channel capacity.
\item Low latency and forwarding cost at routers, switches and protocol transducers.
\item Resilience: if something breaks there is a failover option.
\item Human comprehensibility favors centralization.
\end{itemize}
These are all systemic promises of an architecture. They
enable emergent features of a network, and cannot
be controlled from any single point.

\section{Conclusions}

Promise Theory describes service-oriented functionality, while avoiding
imperative and centralized thinking.  We use it here make neutral assessments of
existing networking principles, and lay bare the functional and
emergent aspects.  We hope that, by summarizing networking in this
manner, the challenges and solutions appear plainer, and ultimately tractable.
This is an unusual viewpoint for Computer Science. Rather than
adopting the conventional belief that only that which is programmed
happens, it takes the opposite viewpoint: ``only that which is
promised can be predicted''. It therefore approaches management by
embracing uncertainty---one might say with realism rather than
faith\cite{burgess_search_2013}.

Some of the challenges we've addressed include:
\begin{itemize}\small
\item The role of trust in winning predictability rather than `control'.

\item Communicating through meaningful abstractions in the context
of applications.

\item Labelling and isolating resources used
by different application owners.
\end{itemize}

How one chooses to solve these is open for discussion. The structure
of promise theory suggests an approach based more on decentralized
abstraction. Bottom-up design favors stability over functionality, but
purpose drives change from the top-down. Striking this balance is a
task for new container abstractions and simplifying patterns.

What we believe the network industry has done well in the past is to
build network infrastructure on a desired end-state model with a low
level of programmatic reasoning.  There is no programming needed to
run OSPF or VLAN for instance, only a few fixed data-driven promises. This limits complexity.
Where it has become deficient is in failing to deal with promises that
aid the running of applications, forcing business users to confront
irrelevant details.

Complexity is a major part of the challenge facing the industry.  To
some extent this is a cognitive issue rather than a fundamental
limitation.  How shall we deal with that? By returning to ``easy''
ideas like centralization and brute force that scale poorly we might
be able to stave off dealing with the issues, but at what cost to
resilience and fault tolerance?

There are many topics we have not been able to cover in this paper.  A
full analysis of current datacenter patterns in relation to different
kinds of application architectures, scales and economies also seems
overdue. Today we hear mainly about the massive social media
datacenters. While this might be relevant for shared cloud services,
there are still questions to be answered about how best to scale
physical networks for different kinds of applications.

It would be interesting to compare and contrast the different approaches
developed by ourselves (the authors) in regard to these challenges, and see how
they could be sewn into a unified view. How does IPv6 fare in the current
picture, what about MPLS and other technologies? How shall we identify
fitness for purpose without losing comprehensibility of networks?

Because of cloud patterns and the resource reusability they enable, it
has become common to say that datacenter traffic patterns are changing
from traditional ``North-South'' to ``East-West'' due to the nature of
compute virtualization. 
Many authors have commented that one could learn from
biological systems in this regard\cite{mark_burgess_computer_1998,gerald_jay_sussman_building_2007},
because biology has clearly evolved systems that embody the properties
we find desirable for building robust networks.  Paul Baran foresaw an
Internet that was a lattice-like mesh\cite{baran} for resilience.
However, the Internet has a hierarchical structure\cite{linked} that
is more fragile, and we tend to favor tidy hierarchies over robust
meshes.  Robustness might appear complex, but its underlying principles
are simple, and can be modelled in our promise-theory based approach.

The challenges of scaling data communications are amongst the most
difficult we now face in IT.  We believe that these are crucial and
exciting times, especially for network innovation, as all aspects of
infrastructure become key players in the design of society's 
critical information
systems.

\section*{Acknowledgments}

We would like to thank Dinesh Dutt for illuminating conversations.
MB would like to thank John Willis.

\bibliographystyle{unsrt}
\bibliography{Promise_Networks}

\end{document}